\documentclass[
reprint,
superscriptaddress,
amsmath,amssymb,
prl
]{revtex4-2}

\usepackage{graphicx}
\usepackage{dcolumn}
\usepackage{bm}
\usepackage{hyperref}
\usepackage[dvipsnames]{xcolor}
\usepackage{soul}
\usepackage{ulem}
\usepackage[left]{lineno}
\usepackage{braket}

\definecolor{dark-green}{RGB}{0, 128, 0}

\begin{document}
\title{Observation of extrinsic topological phases in Floquet photonic lattices}

\author{Rajesh Asapanna}
\thanks{These authors contributed equally.}
\affiliation{Univ. Lille, CNRS, UMR 8523 -- PhLAM -- Physique des Lasers Atomes et Mol\'ecules, F-59000 Lille, France}

\author{Rabih~El~Sokhen}
\thanks{These authors contributed equally.}
\affiliation{Univ. Lille, CNRS, UMR 8523 -- PhLAM -- Physique des Lasers Atomes et Mol\'ecules, F-59000 Lille, France}

\author{Albert~F.~Adiyatullin}
\altaffiliation[Present address: ]{Quandela, 7 Rue Léonard de Vinci, 91300 Massy, France}
\affiliation{Univ. Lille, CNRS, UMR 8523 -- PhLAM -- Physique des Lasers Atomes et Mol\'ecules, F-59000 Lille, France}

\author{Clément~Hainaut}
\affiliation{Univ. Lille, CNRS, UMR 8523 -- PhLAM -- Physique des Lasers Atomes et Mol\'ecules, F-59000 Lille, France}

\author{Pierre~Delplace}
\affiliation{ENS de Lyon, CNRS, Laboratoire de Physique (UMR CNRS 5672), F-69342 Lyon, France}

\author{\'Alvaro~G\'omez-Le\'on}
\email{a.gomez.leon@csic.es}
\affiliation{Institute of Fundamental Physics IFF-CSIC, Calle Serrano 113b, 28006 Madrid, Spain}

\author{Alberto~Amo}
\email{alberto.amo-garcia@univ-lille.fr}
\affiliation{Univ. Lille, CNRS, UMR 8523 -- PhLAM -- Physique des Lasers Atomes et Mol\'ecules, F-59000 Lille, France}

\date{\today}

\begin{abstract}
Discrete-step walks describe the dynamics of particles in a lattice subject to hopping or splitting events at discrete times. 
Despite being of primordial interest to the physics of quantum walks, the topological properties arising from their discrete-step nature have been hardly explored. 
Here we report the observation of topological phases unique to discrete-step walks. 
We use light pulses in a double-fibre ring setup whose dynamics maps into a two-dimensional lattice subject to discrete splitting events.
We show that the number of edge states is not simply described by the bulk invariants of the lattice (i.e., the Chern number and the Floquet winding number) as would be the case in static lattices and in lattices subject to smooth modulations.
The number of edge states is also determined by a topological invariant associated to the discrete-step unitary operators acting at the edges of the lattice.
This situation goes beyond the usual bulk-edge correspondence and allows manipulating the number of edge states without the need to go through a gap closing transition. Our work opens new perspectives for the engineering of topological modes for particles subject to quantum walks.

\end{abstract}

\maketitle

The topological classification of phases of matter is a powerful tool to understand and predict the existence of surface transport channels in a wide a variety of electronic~\cite{Hasan2010}, photonic~\cite{Ozawa2019}, acoustic~\cite{xue_topological_2022} and soft condensed matter systems~\cite{Delplace2017}. 
It relies on the definition of appropriate topological invariants describing global properties of the eigenmode spectrum of an infinite crystalline system. 
In static lattices the most spectacular manifestations of non-trivial topology include the integer and fractional quantum Hall effects, topological insulators and Weyl semimetals, which appear in electronic systems characterised by either Chern or $\mathcal{Z}_2$ indices.
Interestingly, when a lattice is subject to a time-periodic driving described by a time-dependent Hamiltonian, new topological phases appear even when the Chern bulk invariant is zero. 
These are the so-called \textit{anomalous} Floquet topological phases, and they are characterised by a winding number that accounts for the micromotion evolution of the lattice within one driving period~\cite{Kitagawa2012, Rudner2013}. 

In both Chern and anomalous phases, the bulk invariants are computed from the structure of the eigenmodes of an infinitely large system, and their values faithfully describe the number of surface states appearing when the lattice is ended at a boundary. 
However, there is an important class of systems that cannot be described by Hamiltonian operators and whose topological properties remain largely unexplored. 
We refer to systems governed by a discrete-step temporal evolution. 
Examples include active matter in lattices of circulators~\cite{Souslov2017}, light pulses in lattices of coupled ring resonators~\cite{Liang2013,afzal_realization_2020}, acoustic excitations in resonator lattices~\cite{Khanikaev2015} and, importantly, quantum walks in a wide variety of quantum systems such as photons, atoms and ions~\cite{zhang_demonstration_2007, karski_quantum_2009, schmitz_quantum_2009, schreiber_photons_2010, cardano_quantum_2015}, including the boson sampling problem in which indistinguishable photons are subject to a cascade of beamsplitters~\cite{Brod2019}.

In all these systems, the movement of particles towards subsequent sites is described by a series of discrete-step unitary evolution operators. 
The discrete nature of the evolution prevents the characterisation of the dynamics using a continuous microscopic time-dependent Hamiltonian. 
For this reason, the topological toolbox employed to characterise the usual Chern and anomalous phases cannot be applied to systems subject to a discrete-step evolution, and bulk invariants are not sufficient to describe their topological properties~\cite{delplace_phase_2017, cedzich_complete_2018, nitsche_eigenvalue_2019,delplace_topological_2020}

Actually, discrete-step lattices in one and two dimensions have been shown to host anomalous topological edge states similar to those expected in Hamiltonian systems under continuous periodic driving~\cite{Kitagawa2012, Rudner2013, Maczewsky2017, Mukherjee2017a, Bisianov2019}.
But, recently, it has been proposed the existence of topological phases unique to lattices subject to discrete-step dynamics in two dimensions~\cite{bessho_nielsen-ninomiya_2021, bessho_extrinsic_2022}. 
In this case, the number of edge states is given by the combination of a bulk topological invariant and an invariant associated to the winding of the unitary operators acting on the edge.
For this reason this topological phase is referred to as "extrinsic topology", in contrast to the bulk topology that defines the properties of Hamiltonian systems.
Such winding of the edge operators cannot be engineered in conventional Floquet Hamiltonian systems due to the algebraic constraints imposed by the hermiticity of the underlying evolution, but it is possible in systems described by discrete unitary operators, which do not need to be hermitian.

\begin{figure*}[t!]
\includegraphics[width=\textwidth]{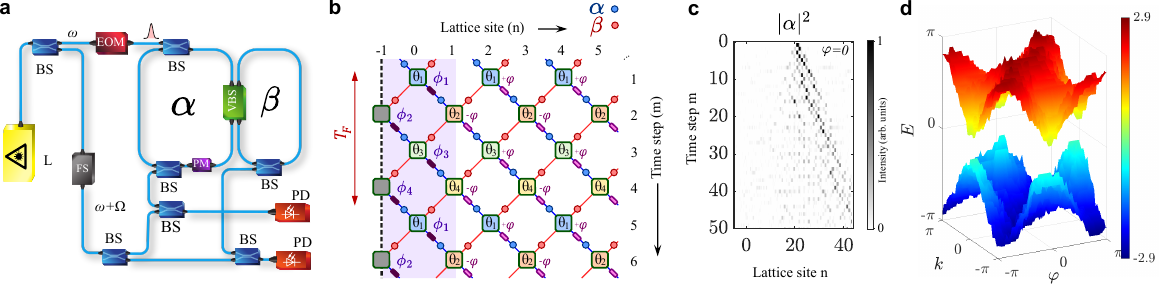}
\caption{\label{fig1} 
\textbf{Discrete-step lattice}. \textbf{a} Scheme of the experimental set-up with beamsplitters BS, variable beamsplitter VBS, electrooptic modulator EOM, phase modulator PM, photodiodes PD and frequency shifter FS to create a local oscilator for the measurement of the eigenvectors and eigenvalues. The $\alpha$ and $\beta$ rings have a length of 45.34m and 44.79m, respectively. 
\textbf{b}~Discrete-step lattice after time demultiplexing of the pulses in the double ring where gray squares correspond to $\theta=\pi/2$ (fully reflecting edge). 
\textbf{c}~Measured light intensity in the $\alpha$ ring in an example of step-evolution after injection at a single site for a lattice model with $\theta_1=0.125\pi$, $\theta_2=0.25\pi$, $\theta_3=0.375\pi$ and $\theta_4=0.125\pi$ and $\varphi=0$. 
\textbf{d}~Measured band tomography as a function of quasimomenta $k$ and $\varphi$ for the $\theta_i$ parameters of~\textbf{c}.
}
\end{figure*}

In this work we provide an experimental implementation of extrinsic topological phases in two dimensions using a synthetic photonic lattice in a system of two-coupled fibre rings filled with coherent pulses of light. 
We show finite-size lattices with a topologically trivial bulk and 0, 1 and 2 edge state bands depending on the winding of the edge operators. 
The discrete-step nature of the system allows for the number of edge states to be modified without passing through a bandgap closing transition.
When a lattice with non-zero Chern number is used, we demonstrate the cancellation of the Chern edge states via the proper design of the step-evolution operators acting on the edge. 
Our results unveil the topological phases that are relevant to quantum walks. 
They show that the engineering of edge states can be performed by acting only on edge sites.

The synthetic lattice system we employ is sketched in Fig.~\ref{fig1}\textbf{a} and described in detail in Refs.~\cite{Adiyatullin2023, el_sokhen_edge-dependent_2024,Supplementary}. 
It is made of two coupled fibre rings of slightly different length. 
A short laser pulse of $\sim1.4$~ns with a large number of photons is injected in the $\alpha$ ring and starts circulating. 
A splitting event takes place at the variable beamsplitter at every round trip generating new pulses that fill discrete time positions $n$ in both the $\alpha$ and $\beta$ rings. 
To compensate for extraction and impedance mismatch losses, we use commercial Er doped amplifiers inside each of the rings.
The dynamics can be mapped into the discrete-step evolution of light pulses in a one-dimensional lattice subject to a discrete-step walk every round trip (see Fig.~\ref{fig1}\textbf{b}, lattice sites are labelled $n$ and time steps $m$).
Similar lattice arrangements have been used to investigate parity-time defects and solitons~\cite{Regensburger2013,Wimmer2015a,Muniz2019}, Bloch oscillations~\cite{Wimmer2015}, anomalous transport~\cite{Wimmer2017}, artificial gauge fields~\cite{Chalabi2019}, the non-hermitian skin effect~\cite{Weidemann2020}, superfluidity~\cite{Wimmer2021} and winding bands~\cite{Adiyatullin2023}.
To study effective two-dimensional systems we introduce a phase modulator that adds a controlled phase $\varphi_m$ to pulses travelling through the $\alpha$ ring with a value alternating between $\varphi$ and $-\varphi$ at odd and even steps. 
The phase $\varphi$ acts as a generalised quasimomentum resulting in a parametric dimension ($\varphi \in (-\pi, \pi]$) in addition to the quasimomentum $k$ (the conjugate of the spatial position $n$ of the pulses in the lattice).

The time evolution of light pulses within the rings follows the set of equations~\cite{Regensburger2011, Bisianov2019, Adiyatullin2023}:

\begin{eqnarray} \label{Eq:step}
\alpha_n^{m+1} &=& \left(\cos\theta_m\alpha_{n-1}^m + i\sin\theta_m\beta_{n-1}^m\right) e^{i\varphi_m} 
\nonumber \\
\beta_n^{m+1} &=& i\sin\theta_m\alpha_{n+1}^m + \cos\theta_m\beta_{n+1}^m
\end{eqnarray}

\noindent with $\alpha_n^{m}$ and $\beta_n^{m}$ being the amplitude of the pulses in the long and short rings, respectively, at spatial position $n$ and time step $m$. 
The splitting amplitude at the variable beamsplitter is $\cos\theta_m$, which can be controlled electronically from fully reflective ($\theta_m=0$) to fully transmissive ($\theta_m=\pi/2$).

Equation (\ref{Eq:step}) can be written in the form of a series of unitary operators $U_m$ acting on an initial vector state: $\ket{\alpha, \beta}_m=U_{m}U_{m-1}\cdots U_1\ket{\alpha, \beta}_0$). 
We employ a sequence of four ordered values of the splitting angle $\theta_1$, $\theta_2$, $\theta_3$, $\theta_4$ which we repeat every four steps (i.e., the Floquet period, see Fig.~\ref{fig1}\textbf{a}): $U_{F}=U_{4}U_{3}U_{2}U_{1}$. 
When periodic boundary conditions are considered, the eigenvalues of $U_{F}$ form two bands separated by gaps at $E=0$ and $E=\pi$ as depicted in Fig.~\ref{fig1}\textbf{d} and derived analytically in Ref.~\cite{Supplementary}.
The periodic Floquet modulation of the lattice results in a Brillouin zone periodic both in quasimomenta and quasienergy.
A typical experimental spatiotemporal diagram of the step evolution in the lattice after injection at a bulk site in the $\alpha$ ring for $\varphi=0$ is shown in Fig.~\ref{fig1}\textbf{c}. 
A heterodyne detection scheme~\cite{Supplementary} allows accessing the relative phase of the pulses at different positions and time steps. 
A direct Fourier transform of the measured spatiotemporal dynamics for a given value of $\varphi$ at Floquet stroboscopic times (every four steps) provides direct experimental access to the eigenvalues of $U_F$ and its eigenvectors~\cite{Adiyatullin2023, el_sokhen_edge-dependent_2024}.
By repeating the experiment at different values of $\varphi$ we can reconstruct the two dimensional bands (Fig.~\ref{fig1}\textbf{d}) and eigenvectors.

\begin{figure}[t!]
\includegraphics[width=\columnwidth]{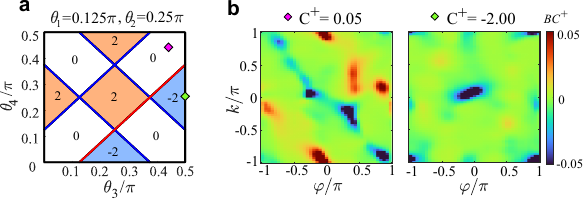}
\caption{\label{fig2} 
\textbf{Chern phase diagram}.
\textbf{a} Bulk topological phases according to the Chern number of the upper band for $\theta_1=0.125\pi$ and $\theta_2=0.25\pi$ as a function of $\theta_3$ and $\theta_4$. Solid lines show the closing of the $E=0$ (blue) and $E=\pi$ (red) gaps. \textbf{b,c} Measured Berry curvature of the upper band for lattice models corresponding to the pink and green diamonds in \textbf{a}, respectively. The measured Chern number is shown on top.}
\end{figure}

Figure~\ref{fig2}\textbf{a} shows the phase diagram of the four steps Floquet model. 
To reduce the number of parameters, we focus on a corner of parameter space in which $\theta_1=0.125\pi$, $\theta_2=0.25\pi$, and $\theta_3$ and $\theta_4$ can vary between $0$ and $0.5\pi$. 
The red lines in the diagram correspond to the closure of the $E=\pi$ gap, and the blue lines of the $E=0$ gap. 
The colours represent the Chern number obtained analytically for the highest energy band.
Figure~\ref{fig2}\textbf{b} and \ref{fig2}\textbf{c} display the measured Berry curvature for the lowest band in the regions marked with diamonds in \textbf{a}.
The Berry curvature is computed from the measured eigenvectors following the procedure developed in Refs.~\cite{blanco_de_paz_tutorial_2020, el_sokhen_edge-dependent_2024} and detailed for this specific case in Ref.~\cite{Supplementary}.
The integration of the experimental Berry curvature over the whole Brillouin zone matches well the expected Chern numbers.

Once we have characterised the topological properties of the bulk bands we draw our attention to the existence of edge states.
In a discrete-step lattice, gapless edge states have two origins~\cite{bessho_extrinsic_2022}.
The first one is the bulk topology of the stroboscopic Floquet operator $U_F$.
According to the Autler-Zinbauer symmetry classification, our two-dimensional lattice is in the D-class with particle-hole symmetry (see Ref.~\cite{Supplementary} for a demonstration). 
Therefore, the bulk topological invariant is the Chern number~\cite{Schnyder2008}.
Contrary to lattice systems described by a microscopic  Hamiltonian, this invariant is not enough to infer the number of edge states in discrete-step walks.
In addition, we need to consider the topology of the unitary operators acting on the considered edge as recently discussed by Bessho and coworkers~\cite{bessho_extrinsic_2022}.
To do so, we separate the unitary Floquet operator $X_{F,N}$ acting on the finite size lattice shown in Fig.~\ref{fig1}\textbf{b} in two parts: $X_{F,N} = U_\text{Edge} U_{F,N}$, where $N$ denotes the number of lattice sites, $U_{F,N}$ is the bulk Floquet operator of the finite size lattice. It is the finite lattice version in real space of $U_F$. The unitary operator $U_\text{Edge}$ describes local modifications at the sites corresponding to one of the edges.
If no particular change of splitting ratios $\theta$ or phase modulator phases $\varphi$ is introduced at the edge sites, $U_\text{Edge}=I$.

The total number of edge states $\mathcal{N}$ at a particular gap and the sign of their group velocity is then given by $\mathcal{N}=\mathcal{C} +\nu_\text{Edge}$, where $\mathcal{C}$ is the Chern number of the band below the gap and $\nu_\text{Edge}$ is the winding of the unitary operator $U_\text{Edge}$. 
In our case we consider an edge in the spatial dimension $n$ (Fig.~\ref{fig1}\textbf{b}), and the winding of the edge operators is defined along the $\varphi$ quasimomentum direction:
\begin{equation}
\nu_\text{Edge}= \frac{1}{2\pi}\int_0^{2\pi} d\varphi \ \text{Tr}[U_\text{Edge}(\varphi)^{-1} i \partial_\varphi U_\text{Edge}(\varphi)].
\end{equation}

\begin{figure}[t!]
\includegraphics[width=\columnwidth]{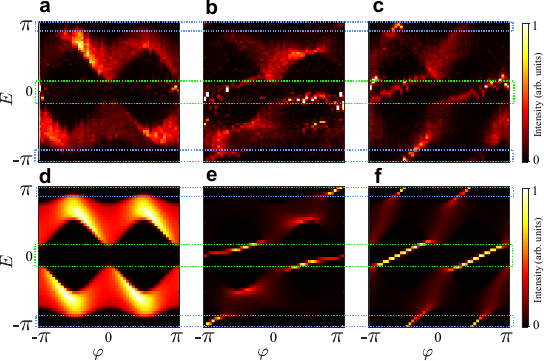}
\caption{\label{fig3}
\textbf{Extrinsic topology with $\mathcal{C}=0$}.
\textbf{a} Measured band diagram (intensity in the $\alpha$ ring) for a model with $\theta_1=0.125\pi$, $\theta_2=0.25\pi$, $\theta_3=0.438\pi$ and $\theta_4=0.438\pi$ and full reflection at the boundary in site $n=-1$. Injection takes place at $n=0$. 
\textbf{b,c} Same as \textbf{a} when phase modulators add a winding of $-1$ and $-2$, respectively, to the edge operators. The colour scale is lightly saturated for a better visualisation of the edge state bands. 
\textbf{d,e,f} Numerical simulations of Eq.\ref{Eq:step} in the conditions of \textbf{a,b,c}, respectively. 
In the band diagrams, the dotted green box highlights the $E=0$ gap and the dotted blue boxes combined highlight the $E=\pi$ gap.}
\end{figure}

Recently, we showed that the geometry of the edges plays an important role in the existence of edge states in discrete-step lattices~\cite{el_sokhen_edge-dependent_2024}. 
Let us now demonstrate that the number of edge states in the  lattice under consideration can be manipulated by modifying the winding of the edge unitary operators. 
We implement a lattice with an edge with the void in the spatial dimension by setting the splitting angle $\theta$ to $\pi/2$ at the leftmost splitter, as illustrated in Fig.~\ref{fig1}\textbf{b}.
Figure~\ref{fig3}\textbf{a} displays the measured band diagram in the region with zero Chern number corresponding to the pink diamond in Fig.~\ref{fig2}\textbf{a} when an initial pulse is injected at site 0 in the $\alpha$ ring.
Two bands are separated by two gaps at $E=0$ and $E=\pi$, and their shape and population match the simulation of Eq.~(\ref{Eq:step}) in the same conditions (Fig.~\ref{fig3}\textbf{d}).
No edge state bands traverse any of the gaps as expected from a phase with $\mathcal{C}=0$ and no winding of the edge operators: $U_\text{Edge}=I$.

We can induce the emergence of edge state bands traversing both gaps by designing edge operators with $\nu_\text{Edge}\neq 0$.
To do so, we modify the values of the phase modulator at the edge sites (marked in violet in Fig.~\ref{fig1}\textbf{b}) to follow the sequence $\phi_1=\varphi$, $\phi_2=0$, $\phi_3=-\varphi$ and $\phi_4=\varphi$ at each Floquet period. 
The phase modulators in the bulk are not modified and keep the sequence $+\varphi$, $-\varphi$ at odd and even steps.
A single band of edge states traverses both gaps as seen in Fig.~\ref{fig3}\textbf{b} in the experiments and Fig.~\ref{fig3}\textbf{e} in the simulations.
These states are localised at the left edge of the lattice, see Ref.~\cite{Supplementary}.
The reason for their appearance is that the edge unitary operator $U_\text{Edge}$ now has a non-trivial winding given by $\nu_\text{Edge}=-\sum_{i=0}^4 \boldsymbol{c_i}$ with $\boldsymbol{c_i}=\phi_i/\varphi$ (see Ref.~\cite{Supplementary}). In this case, $\nu_\text{Edge}=-1$.
Therefore, any number edge state bands with desired group velocity can be implemented. 
For instance, we can increase the number of chiral edge states by increasing the winding of the edge operators using the sequence $\phi_1=\phi_3=\varphi$, $\phi_2=\phi_4=0$ as illustrated in Fig.~\ref{fig3}\textbf{c} and Fig.~\ref{fig3}\textbf{f}, now two edge state bands traverse each gap with $\nu_\text{Edge}=-2$.

\begin{figure}[t!]
\includegraphics[width=\columnwidth]{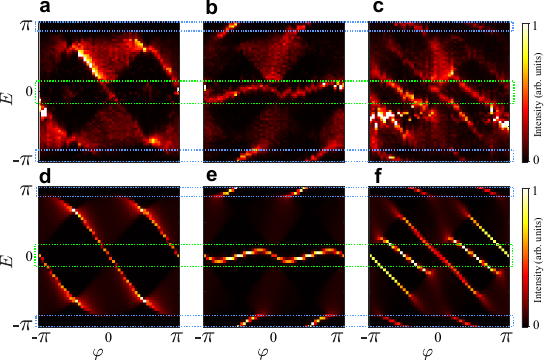}
4\caption{\label{fig4}
\textbf{Extrinsic topology with }$\mathcal{C}=-2$ for upper band ($\mathcal{C}=+2$ for lower band).
\textbf{a} Measured band diagram (intensity of the $\alpha$ ring) for  a model with $\theta_1=0.125\pi$, $\theta_2=0.25\pi$, $\theta_3=0.5\pi$ and $\theta_4=0.25\pi$ and full reflection at the boundary in site $n=-1$. 
Injection takes place at $n=0$. 
\textbf{b,c} Same as \textbf{a} when phase modulators add a winding of $-2$ and $+2$, respectively, to the edge operators. 
The colour scale is lightly saturated for a better visualisation of the edge state bands.
\textbf{d,e,f} Numerical simulations of Eq.\ref{Eq:step} in the conditions of \textbf{a,b,c}, respectively. 
In the band diagrams, the dotted green box highlights the $E=0$ gap and the dotted blue boxes combined highlight the $E=\pi$ gap.}
\end{figure}

An interesting situation occurs when the lattice has non-trivial Chern number.
This is studied in Fig.~\ref{fig4} for a lattice in the blue region marked with a green diamond in Fig.~\ref{fig2}\textbf{a}, with $\mathcal{C}=-2$ for upper band ($\mathcal{C}=+2$ for lower band).
When the operators at the edge follow the $+\varphi$, $-\varphi$ sequence of the bulk, the measured and simulated band diagrams in Fig.~\ref{fig4}\textbf{a} and \textbf{d}, respectively, display two bands of edge states traversing the $E=0$ gap, and no edge states traversing the $E=\pi$ gap.
This is what we would expect from a pure Chern phase with $\nu_\text{Edge}=0$.
The extrinsic topology can cancel or enrich the number of edge states of Chern origin.
Figures~\ref{fig4}\textbf{b} and \textbf{e} show the case of $\nu_\text{Edge}=-2$, with the set of edge phase modulators $\phi_1=\phi_3=\varphi$, $\phi_2=\phi_4=0$.
The winding of the edge unitary operators cancels the Chern edge states resulting in a trivial edge mode at the $E=0$ gap that wiggles but does not connect the bands. 
Simultaneously, two new edge state bands appear at the $E=\pi$ gap.
If $\nu_\text{Edge}$ is modified to have the value of $+2$ through the sequence $\phi_1=\phi_3=-\varphi$, $\phi_2=\phi_4=0$, the middle gap now shows four edge state bands.
They are the addition of the Chern and extrinsic topology edge states.

\begin{figure}[t!]
\includegraphics[width=0.75\columnwidth]{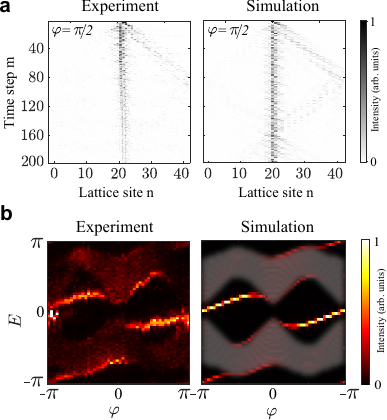}
\caption{\label{fig5} 
\textbf{Winding modes localised in the bulk}. \textbf{a} Measured (left) and simulated (right) spatiotemporal evolution in the $\alpha$ sites in a lattice with $\mathcal{C}=0$ ($\theta_i$ values of Fig.~\ref{fig3}) when a pulse is injected at a bulk site $n_b=22$ in which a winding of -1 has been implemented ($\phi_1^{n_b}=\varphi$, $\phi_2^{n_b}=\phi_3^{n_b}=\phi_4^{n_b}=0$), with $\varphi=\pi/2$.
\textbf{b} Band diagram as a function of $\varphi$. A mode traverses both gaps.
The semi-transparent overlay in the right panel highlights the bulk modes.
}
\end{figure}

Finally, we take advantage of the possibility to induce non-trivial windings in local unitary operators to arrange chiral modes within the bulk of the lattice.
For this purpose we use a lattice with parameters corresponding to the pink diamond in Fig.~\ref{fig2}\textbf{a} with bands with $\mathcal{C}=0$. 
The bulk bands are displayed in Fig.~\ref{fig3}\textbf{a} and \textbf{d}.
To create chiral states within the bulk, we select a cell far from the edges ($n_b=22$) and implement locally the Floquet sequence of phases $\phi_1^{n_b}=\varphi$, $\phi_2^{n_b}=\phi_3^{n_b}=\phi_4^{n_b}=0$.
The winding of the local $U_{n_{b}}$ operators is -1.
If we inject a pulse of light in the $\alpha$ ring at this site, we observe a mode bound to this unit cell in Fig.~\ref{fig5}\textbf{a}. 
The observed residual spread out of the mode arises from the coupling of the initial injection to bulk modes.
A tomography of the bands as a function of the $\varphi$ quasimomentum is displayed in Fig.~\ref{fig5}\textbf{b}. 
It reveals a single band of states traversing each of the gaps once.
They are unidirectional chiral in-gap modes in the bulk of an otherwise homogeneous lattice.
Similar results were found in the bulk of lattices with $\mathcal{C}=\pm2$ (not shown).
Therefore, their origin is not related to a change of topology at either side of an interface, but to the winding of the discrete-step unitary operators acting on a particular site.
Similarly to the extrinsic edge modes, these bulk localised modes can be described via the winding of a local operator~\cite{Supplementary}.

Even though the results reported in this work focus on light pulses in the classical regime, they provide a crucial demonstration of topological phases relevant for quantum walks in which individual particles are subject to discrete-step dynamics. These include Fock and other few photon/boson states.
Our work calls for further studies on the effects of quantum interference in this type of setting. 
In the classical regime an exciting perspective is the investigation of nonlinear effects.
Kerr-type nonlinearities, which modify the phase of a wavepacket, could result in nonlinear windings in such lattices.

\textit{Acknowledgements}
We thank Martin Guillot for his help in the implementation of averaging methods to extract the Berry curvature. This work was supported by the European Research Council grant EmergenTopo (865151), the French government through the Programme Investissement d'Avenir (I-SITE ULNE /ANR-16-IDEX-0004 ULNE) managed by the Agence Nationale de la Recherche, the Labex CEMPI (ANR-11-LABX-0007) and the region Hauts-de-France. This project has received funding from the European Union’s Horizon 2020 research and innovation programme under the Marie Skłodowska-Curie grant agreement No 101108433.
A.G.L acknowledges support by MICIU/AEI/10.13039/501100011033 and by ERDF/EU to project PID2023-146531NA-I00. Also acknowledges support from CSIC Interdisciplinary Thematic Platform (PTI+) on Quantum Technologies (PTI-QTEP+).


\end{document}